\documentclass[letterpaper, 10 pt, conference]{ieeeconf}  

\IEEEoverridecommandlockouts                              

\usepackage{graphics} 
\usepackage{epsfig} 
\usepackage{mathptmx} 
\usepackage{times} 
\usepackage{amsmath} 
\usepackage{amssymb}  
\usepackage{booktabs}
\usepackage{tabularx}
\title{\LARGE \bf
Route-Align-Verify for Functional Correctness in Code Generation
}

\author{Erxue Zhou$^{1}$ \quad Jingxiang Meng$^{2}$ \quad Aofan Liu$^{3}$%
\thanks{$^{1}$Erxue Zhou is with the Software Engineering Institute, East China Normal University, Shanghai 200062, China
        {\tt\small 51275902066@stu.ecnu.edu.cn}}%
\thanks{$^{2}$University of Chicago.}%
\thanks{$^{3}$Peking University.}%
}

\begin{document}

\maketitle
\thispagestyle{empty}
\pagestyle{empty}

\begin{abstract}
Large language models (LLMs) have substantially improved code generation, yet achieving strong \textbf{functional correctness} remains difficult, especially on heterogeneous programming tasks where a single prompting strategy and a single direct output are often insufficient. In this paper, we present \textbf{RAV}, a lightweight and modular framework that improves code generation under a fixed backbone model through three coordinated stages: \textbf{Route}, which applies task-aware prompt routing before generation; \textbf{Align}, which reduces the mismatch between fine-tuning prompts and inference-time prompts via aligned LoRA adaptation; and \textbf{Verify}, which selects final outputs by executing multiple candidates against visible public tests. 

We evaluate RAV on the MBPP benchmark under both \texttt{sanitized} and \texttt{full} settings. The full RAV pipeline achieves the best results among all evaluated configurations, reaching \textbf{0.8911} on MBPP Sanitized and \textbf{0.8520} on MBPP Full. Compared with the base model, this corresponds to improvements of \textbf{6.35} and \textbf{9.92} percentage points, respectively. A component-wise ablation further shows that task-aware routing and aligned adaptation become substantially more effective when coupled with execution-based verification. Additional robustness and contamination analyses support the reliability of the observed gains. Overall, the results suggest that meaningful improvements in code generation functional correctness can be obtained without modifying the backbone architecture itself, by jointly optimizing how tasks are prompted, how the model is adapted, and how final outputs are selected.
\end{abstract}
\section{Introduction}

Large language models (LLMs) have become increasingly capable of synthesizing executable code from natural-language problem descriptions, docstrings, and partial implementations \cite{chen2021codex,austin2021program,li2023starcoder,roziere2023codellama}. This progress has been accompanied by execution-based benchmarks such as HumanEval and MBPP, which evaluate generated programs by whether they actually pass tests rather than merely appearing plausible at the surface level \cite{chen2021codex,austin2021program}. Broader benchmarks such as APPS, stronger evaluation suites such as EvalPlus, and more recent contamination-aware benchmarks such as LiveCodeBench further highlight that functional correctness, rather than textual plausibility alone, is a central criterion for reliable code generation \cite{hendrycks2021apps,liu2023evalplus,jain2024livecodebench}. Despite rapid progress, achieving strong functional correctness remains difficult, especially on heterogeneous tasks that differ substantially in style, structure, and failure modes.

A central challenge is that code-generation performance depends on more than the backbone model alone. First, different tasks often respond differently to prompting style: a single generic prompt may be suboptimal for string manipulation, arithmetic reasoning, or edge-case-heavy problems \cite{austin2021program,wizardcoder2023,guo2024deepseekcoder}. Second, the effectiveness of adaptation is influenced by how training instructions are constructed and matched to downstream use, so a mismatch between fine-tuning prompts and inference-time prompts can weaken the benefit of specialization \cite{wizardcoder2023,magicoder2023,guo2024deepseekcoder}. Third, even when a model is capable of producing a correct solution, that solution may not appear in the first sample, making one-shot decoding an inefficient way to exploit the model's latent capability \cite{chen2021codex,li2022alphacode,chen2022codet}. These observations suggest that improving functional correctness may require coordinated optimization across \textit{prompting}, \textit{adaptation}, and \textit{selection}, rather than relying on only one of them.

Prior work has explored these directions largely in isolation. Code-specialized pretraining and instruction tuning improve code generation by eliciting behaviors better matched to downstream tasks, as shown by systems such as WizardCoder, Magicoder, StarCoder, and Code Llama \cite{wizardcoder2023,magicoder2023,li2023starcoder,roziere2023codellama}. Parameter-efficient adaptation methods such as LoRA make such specialization practical without updating all backbone parameters \cite{hu2021lora}. At test time, execution-aware methods such as CodeT and Self-Debug show that generating multiple candidates and leveraging execution signals can substantially improve final correctness \cite{chen2022codet,chen2023selfdebug}; related large-scale sampling and filtering strategies have also proved effective in systems such as AlphaCode \cite{li2022alphacode}. However, these directions are typically studied as separate techniques rather than as components of a coordinated pipeline. In particular, the interaction among \textit{task-aware prompting}, \textit{training--inference prompt alignment}, and \textit{lightweight execution-based selection} remains underexplored.

In this paper, we propose \textbf{RAV}, short for \textbf{Route-Align-Verify}, a simple modular framework for improving code generation under a fixed backbone model. The framework consists of three stages. \textbf{Route} applies lightweight task-aware prompt routing before generation. \textbf{Align} fine-tunes the model using prompts rewritten to better match the routed prompt style used at test time. \textbf{Verify} samples multiple candidate programs and selects a final output using execution-based verification on visible public tests. Rather than modifying the backbone architecture itself, RAV aims to improve functional correctness by coordinating how tasks are asked, how the model is adapted, and how outputs are selected.

We evaluate RAV on MBPP under both \texttt{sanitized} and \texttt{full} settings. The full RAV pipeline achieves the best performance among the evaluated configurations, improving pass@1 from 0.8276 to 0.8911 on MBPP Sanitized and from 0.7528 to 0.8520 on MBPP Full. A component-wise ablation further shows that routing and aligned adaptation alone provide only limited direct gains, but become substantially more effective when coupled with verification. This empirical pattern suggests that Route and Align primarily improve the quality of the candidate pool, while Verify converts that advantage into stronger final functional correctness.

The main contributions of this paper are as follows:
\begin{itemize}
    \item We propose \textbf{RAV}, a modular Route-Align-Verify framework that improves code generation functional correctness without modifying the backbone architecture.
    \item We introduce a practical \textbf{training--inference prompt alignment} strategy based on aligned LoRA adaptation, which reduces mismatch between fine-tuning prompts and routed prompts at test time.
    \item We present a \textbf{component-wise empirical study} on MBPP showing that task-aware routing, aligned adaptation, and execution-based verification are most effective when used jointly.
    \item We provide \textbf{robustness and contamination analyses} that support the reliability of the observed gains.
\end{itemize}
\section{Related Work}

\subsection{Code Generation and Functional Correctness Evaluation}
Large language models have shown strong capability in program synthesis and code completion, with early large-scale studies demonstrating that model scale and code-specific pretraining can substantially improve executable code generation \cite{chen2021codex,austin2021program}. Benchmark design has played a central role in this progress. HumanEval introduced execution-based evaluation for docstring-to-function synthesis \cite{chen2021codex}, while MBPP focuses on short Python programming tasks and has become a standard benchmark for measuring functional correctness under test execution \cite{austin2021program}. Beyond these relatively compact benchmarks, APPS broadens the evaluation space to more diverse and challenging natural-language-to-code problems, further highlighting the importance of execution-based assessment for code generation systems \cite{hendrycks2021apps}.

Recent work has also shown that reported performance can be sensitive to the adequacy of benchmark test suites. EvalPlus revisits widely used code-generation benchmarks by augmenting them with substantially stronger tests and shows that limited test coverage can overestimate model capability \cite{liu2023evalplus}. Motivated by this evaluation perspective, our work is positioned squarely in the functional-correctness setting. Rather than introducing a new benchmark or a new backbone architecture, we study how to improve pass@1 on MBPP through a modular generation pipeline.

\subsection{Instruction Tuning and Parameter-Efficient Adaptation for Code LLMs}
A major direction in code LLM research is to improve downstream performance through instruction tuning, synthetic data construction, and code-specialized training. WizardCoder adapts Evol-Instruct to the code domain and shows that instruction tuning can substantially improve code-generation performance across several benchmarks \cite{wizardcoder2023}. Magicoder further demonstrates the effectiveness of synthetic code instruction data built from open-source snippets, narrowing the gap between open-source and stronger proprietary code models \cite{magicoder2023}. More recently, DeepSeek-Coder shows that strong open code models can also benefit from large-scale code pretraining together with instruction-oriented variants for downstream use \cite{guo2024deepseekcoder}. Taken together, these works suggest that code generation quality depends not only on model scale, but also on how training instructions and code data are constructed.

At the same time, parameter-efficient adaptation methods have become a practical way to specialize large models without updating all backbone parameters. LoRA is a particularly influential approach, freezing the pretrained model weights and learning low-rank updates for downstream adaptation \cite{hu2021lora}. QLoRA further improves adaptation efficiency by combining quantization with LoRA, making high-quality fine-tuning feasible under tighter hardware constraints \cite{dettmers2023qlora}. In our setting, we adopt LoRA not merely as a memory-saving fine-tuning tool, but as a mechanism for \textit{prompt alignment}: the adapter is trained on prompts rewritten to better match the routed prompt style later used during inference.

Compared with prior instruction-tuning work, our contribution is narrower but more targeted. We do not propose a new synthetic data pipeline, a new backbone model, or a new adaptation method. Instead, we emphasize the practical importance of reducing the mismatch between fine-tuning prompts and test-time prompts inside a larger code-generation pipeline.

\subsection{Prompting, Verification, and Test-Time Improvement}
Beyond model training, many recent works improve code generation at inference time. Prompt design has been shown to meaningfully affect code generation quality, especially on diverse tasks where a single prompt style may be suboptimal \cite{austin2021program,wizardcoder2023,guo2024deepseekcoder}. In parallel, multi-sample decoding and behavioral filtering exploit the fact that correct solutions may exist among multiple candidates even when the first sample is wrong. Codex reports substantial gains from repeated sampling \cite{chen2021codex}, and AlphaCode further demonstrates the effectiveness of large-scale sampling followed by filtering based on program behavior \cite{li2022alphacode}.

Execution-aware methods make this idea more explicit. CodeT reranks model-generated programs using generated tests, showing that test-based selection can significantly improve final correctness \cite{chen2022codet}. Self-Debug teaches models to debug or revise generated programs using execution feedback, achieving substantial gains on code-generation tasks including MBPP \cite{chen2023selfdebug}. More generally, iterative refinement frameworks such as Self-Refine demonstrate that large models can improve outputs through feedback-driven refinement at test time \cite{madaan2023selfrefine}.

Our Verify stage is most closely related to execution-based selection methods such as CodeT, but it is intentionally lightweight: instead of generating new tests or running a full iterative debugging loop, we use the visible public tests already available in MBPP to rank candidates. More importantly, our work studies verification \textit{together with} task-aware routing and prompt-aligned adaptation. In this sense, RAV differs from prior methods that focus primarily on either prompting, training, or test-time refinement in isolation. Our goal is to show that these stages can interact constructively within a single modular framework.
\section{Method}

\subsection{Problem Setting}
We study code generation under a fixed backbone language model and aim to improve \textbf{functional correctness} on benchmark tasks. Let a programming task be denoted by $x$, where $x$ may include a natural-language problem description, a function signature, a docstring, and visible public tests. Given $x$, the goal is to generate a Python program $c$ that passes the benchmark evaluation.

In this work, we focus on the \textbf{MBPP} benchmark, a collection of short Python programming tasks evaluated under two commonly used settings: \texttt{sanitized} and \texttt{full}. Our main metric is the success rate of the final returned program under benchmark execution. When exactly one final completion is submitted per task, we refer to this quantity as \texttt{pass@1} by convention. We retain this terminology for consistency with prior code-generation evaluation practice, even though some settings first generate multiple candidates and then select one final program.  

We propose \textbf{RAV}, short for \textbf{Route-Align-Verify}, a modular framework that improves code generation at three stages:
\begin{enumerate}
    \item \textbf{Route}: choose a task-aware prompt template before generation;
    \item \textbf{Align}: fine-tune the model using prompts that better match the inference-time routed prompts;
    \item \textbf{Verify}: sample multiple candidate programs and select one using lightweight execution-based verification.
\end{enumerate}

The central hypothesis is that functional correctness can be improved without changing the backbone architecture if the model is (i) prompted in a task-aware way, (ii) adapted to the same prompt style used at test time, and (iii) allowed to select from multiple candidates using inexpensive execution signals.

\subsection{Framework Overview}
\begin{figure*}[t]
\centering
\includegraphics[width=0.95\textwidth]{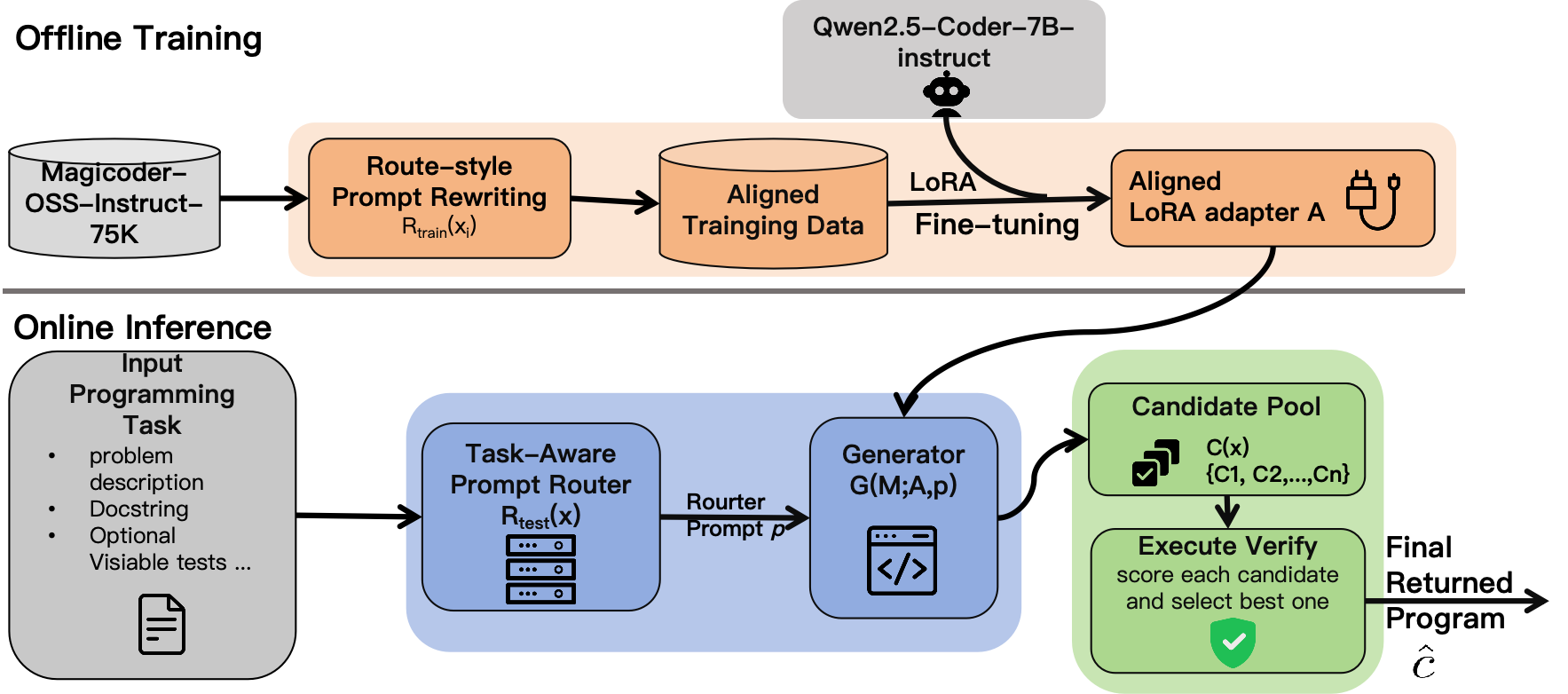}
\caption{Overview of the Route-Align-Verify (RAV) framework. The upper branch shows aligned LoRA fine-tuning with routed-style prompts. The lower branch shows the inference pipeline, including task-aware prompt routing and execution-based candidate selection.}
\label{fig:architecture}
\end{figure*}

The overall architecture of RAV is shown in Fig.~\ref{fig:architecture}. Given an input task $x$, the framework proceeds as follows:
\begin{enumerate}
    \item A router $R_{\text{test}}(\cdot)$ maps the raw task to a routed prompt
    \begin{equation}
        p = R_{\text{test}}(x).
    \end{equation}
    
    \item A generator $G$, parameterized by a backbone model $M$ and an optional aligned LoRA adapter $A$, samples one or more candidate programs:
    \begin{equation}
        \mathcal{C}(x) = \{c_1, c_2, \dots, c_n\}, \qquad c_i \sim G(M; A, p).
    \end{equation}
    
    \item If verification is disabled, the system returns the first sampled candidate:
    \begin{equation}
        \hat{c} = c_1.
    \end{equation}
    
    \item If verification is enabled, a verifier $V$ scores each candidate and selects the best one:
    \begin{equation}
        \hat{c} = \arg\max_{c_i \in \mathcal{C}(x)} V(c_i, x).
    \end{equation}
\end{enumerate}

This definition is important for interpreting the experiments. In particular, settings with \texttt{Verify = off} and $n > 1$ should be understood as \textbf{multi-sample generation without execution-based reranking under a first-sample return policy}, rather than as best-of-$n$ decoding. Under this implementation, different subsets of Route, Align, and Verify can be combined to form a \textbf{component-wise ablation} of the overall framework.

\subsubsection{Module Summary}
\begin{table*}[t]
\centering
\small
\begin{tabular}{@{}llll@{}}
\toprule
\textbf{Module} & \textbf{Stage} & \textbf{Purpose} & \textbf{Main signal} \\
\midrule
Route  & Before generation & Choose task-suitable prompt & Task keywords, docstrings \\
Align  & Training          & Reduce prompt mismatch      & Routed prompt rewriting \\
Verify & After generation  & Select a better candidate   & Public-test execution score \\
\bottomrule
\end{tabular}
\caption{Summary of the three RAV components.}
\label{tab:rav_modules}
\end{table*}

\subsection{Backbone Model and Aligned LoRA}
All experiments use the same backbone model, \textbf{Qwen2.5-Coder-7B-Instruct}. To adapt the model without modifying the full backbone parameters, we adopt \textbf{LoRA} for parameter-efficient supervised fine-tuning. The backbone weights remain frozen while only low-rank update matrices are learned.

In the main setting of this paper, we use an \textbf{aligned LoRA} adapter. The core idea is to reduce the mismatch between the prompts used during fine-tuning and the prompts used at inference time. Rather than training on generic instruction prompts and testing on routed prompts, we rewrite the training inputs into the same routed style expected during inference. This makes the adapter better matched to the actual prompting regime used by the full system.

Let the backbone be $M$ and the aligned LoRA adapter be $A$. We denote the routed generator by
\begin{equation}
    G_{\text{base}}(p) = G(M; \varnothing, p), \qquad
    G_{\text{align}}(p) = G(M; A, p),
\end{equation}
where $\varnothing$ means no adapter is attached.

\subsubsection{Aligned LoRA Training Configuration}
\begin{table}[t]
\centering
\small
\begin{tabular}{@{}ll@{}}
\toprule
\textbf{Item} & \textbf{Setting} \\
\midrule
Base model & Qwen2.5-Coder-7B-Instruct \\
Fine-tuning type & LoRA (rank 16, target \texttt{all}) \\
Cutoff length & 1024 \\
Max samples & 12000 \\
Learning rate & $8\times10^{-5}$ \\
Epochs / Scheduler & 1.0 / cosine \\
Precision & bf16 \\
\bottomrule
\end{tabular}
\caption{Training configuration of the aligned LoRA adapter used in the main experiments.}
\label{tab:aligned_lora_config}
\end{table}

\subsection{Route: Task-Aware Prompt Routing}
A single generic prompt is often insufficient for heterogeneous code-generation tasks. MBPP contains diverse problem types, including string manipulation, arithmetic reasoning, and tasks with explicit corner cases. We therefore introduce a lightweight \textbf{router} that converts each raw task into a more suitable prompt template using task heuristics.

For an input task $x$, the router outputs a prompt template index $r$ and a routed prompt
\begin{equation}
    r = \rho(x), \qquad p = R_{\text{test}}(x; r).
\end{equation}

\subsubsection{MBPP Router}
For MBPP, the router uses lexical and task-structure cues to dispatch tasks into several prompt styles:
\begin{itemize}
    \item \textbf{string\_relaxed}: for string-heavy tasks, triggered by keywords such as \textit{substring}, \textit{palindrome}, and similar text-processing cues;
    \item \textbf{algo\_relaxed}: for algorithmic or math-oriented tasks, triggered by keywords such as \textit{prime}, \textit{gcd}, and other arithmetic patterns;
    \item \textbf{edge\_relaxed}: for tasks with explicit edge-case hints or relatively dense test descriptions;
    \item \textbf{default}: a fallback prompt used when no specialized trigger is activated.
\end{itemize}

The router is intentionally lightweight. Its goal is not to solve the task itself, but to choose a more appropriate prompting style before the model generates code.

\subsection{Align: Training--Inference Prompt Alignment}
The purpose of \textit{Align} is to reduce the distribution gap between fine-tuning prompts and inference-time routed prompts. Let the original supervised training set be
\begin{equation}
    D = \{(x_i, y_i)\}_{i=1}^{N},
\end{equation}
where $x_i$ is the training input and $y_i$ is the reference program. Instead of directly fine-tuning on $x_i$, we construct an aligned dataset
\begin{equation}
    D' = \{(R_{\text{train}}(x_i), y_i)\}_{i=1}^{N},
\end{equation}
where $R_{\text{train}}(\cdot)$ is a training-time rewriting rule designed to approximate the routed prompt style used at inference time.

Conceptually, Align does not add a new reasoning module at test time. Rather, it changes what the adapter has learned to respond to. This makes the generator more compatible with the routed prompts used by Route and improves the quality of the candidate pool supplied to the downstream verifier.

\subsection{Verify: Execution-Based Candidate Selection}
Given a routed prompt $p$, the generator produces a candidate pool
\begin{equation}
    \mathcal{C}(x) = \{c_1, c_2, \dots, c_n\}.
\end{equation}
For MBPP, visible public tests are available together with the task description. We therefore use lightweight execution-based verification to score each candidate against these public tests.

Let $T(x)$ denote the set of visible public tests for task $x$. The verification score of candidate $c_i$ is defined as  
\begin{equation}
    V(c_i, x) = \sum_{t \in T(x)} \mathbf{1}[c_i \text{ passes } t],
\end{equation}
where $\mathbf{1}[\cdot]$ is the indicator function. The final selected program is
\begin{equation}
    \hat{c} = \arg\max_{c_i \in \mathcal{C}(x)} V(c_i, x).
\end{equation}
If ties occur, we apply a weak simplicity prior by preferring shorter code.

This verifier is deliberately lightweight: it does not require formal specification or expensive search. Instead, it uses the public tests already available in MBPP to rank candidates and improve the probability that the final returned program is functionally correct.

\subsection{Complexity and Inference Cost}
Routing adds negligible overhead because it only applies simple task heuristics before generation. Aligned LoRA does not introduce a separate test-time module beyond attaching the learned adapter. The main additional inference cost comes from generating multiple candidates and, when enabled, executing them on public tests.

Let $G$ denote the cost of generating one candidate and $E$ denote the cost of executing one candidate on the public tests. Then the overall test-time cost is
\begin{equation}
    \mathcal{O}\!\left(nG + \mathbf{1}[\text{Verify}] \cdot nE\right),
\end{equation}
where $n$ is the number of sampled candidates. In practice, this makes RAV a modular accuracy--cost tradeoff: better performance can be obtained by verification over a candidate pool, while the underlying model architecture remains unchanged.

\subsection{Summary}
RAV improves code generation through three coordinated interventions: \textbf{Route} changes how tasks are prompted, \textbf{Align} changes how the model is adapted to those prompts, and \textbf{Verify} changes how the final output is selected from generated candidates. Together, these components provide a practical and reproducible way to improve functional correctness on MBPP without modifying the backbone architecture itself.
\section{Experiments}

\subsection{Experimental Setup}

\subsubsection{Benchmark and Metric}
We evaluate the proposed method on the \textbf{MBPP} benchmark under two commonly used settings: \textbf{MBPP Sanitized} and \textbf{MBPP Full}. Performance is measured by \textbf{pass@1}, i.e., the proportion of tasks for which the final returned program passes the benchmark tests. This metric directly reflects functional correctness and is therefore appropriate for evaluating code generation systems.

\subsubsection{Backbone Model and Environment}
All experiments utilize the \textbf{Qwen2.5-Coder-7B-Instruct} backbone. The evaluation was conducted on an NVIDIA GeForce RTX 5090 GPU (32GB VRAM, Driver 580.105.08) using Python 3.11.14 on Linux 5.15.0.

\subsubsection{Compared Configurations}
To analyze the role of each component, we evaluate several representative combinations of \textit{Route}, \textit{Align}, and \textit{Verify}:
\begin{itemize}
    \item \textbf{Base}: the backbone model without routing, aligned adaptation, or verification.
    \item \textbf{Route + Verify}: task-aware routing combined with execution-based verification.
    \item \textbf{Align + Verify}: aligned LoRA combined with execution-based verification.
    \item \textbf{Route + Align}: routing and aligned LoRA without verification.
    \item \textbf{Full RAV}: the complete system with \textbf{Route + Align + Verify}.
\end{itemize}

Since not all individual component toggles are available in the current experiment matrix, we report these results as a \textbf{partial component-wise ablation} rather than a full factorial ablation.

\subsection{Component-wise Ablation on MBPP}

Table~\ref{tab:ablation_mbpp} presents the main results as a component-wise ablation study. The full RAV pipeline achieves the best performance on both MBPP settings, reaching \textbf{0.8911} on MBPP Sanitized and \textbf{0.8520} on MBPP Full.

\begin{table*}[t]
\centering
\caption{Partial ablation study of RAV components on MBPP. $\Delta$ denotes the absolute improvement over the base model on MBPP Full. Best results are shown in bold.}
\label{tab:ablation_mbpp}
\small
\begin{tabular}{lcccccc}
\toprule
Configuration & Route & Align & Verify & Sanitized & Full & $\Delta$ Full \\
\midrule
Base &  &  &  & 0.8276 & 0.7528 & -- \\
Route + Verify & \checkmark &  & \checkmark & 0.8794 & 0.8500 & +0.0972 \\
Align + Verify &  & \checkmark & \checkmark & 0.8833 & 0.8460 & +0.0932 \\
Route + Align & \checkmark & \checkmark &  & 0.8366 & 0.7640 & +0.0112 \\
Full RAV & \checkmark & \checkmark & \checkmark & \textbf{0.8911} & \textbf{0.8520} & \textbf{+0.0992} \\
\bottomrule
\end{tabular}
\end{table*}

Several observations can be drawn from Table~\ref{tab:ablation_mbpp}. First, both \textit{Route + Verify} and \textit{Align + Verify} substantially outperform the base model, indicating that either routing or aligned adaptation can provide a more favorable candidate pool when coupled with execution-based selection. Second, \textit{Route + Align} without verification yields only a modest improvement over the base model, suggesting that better prompting and better adaptation alone do not automatically translate into large pass@1 gains when the system directly returns a sampled candidate. Third, the full RAV configuration consistently exceeds both \textit{Route + Verify} and \textit{Align + Verify}, indicating that Route and Align are complementary under the Verify stage rather than redundant.

Compared with the base model, Full RAV improves pass@1 by \textbf{6.35} points on MBPP Sanitized and by \textbf{9.92} points on MBPP Full. These gains show that the full pipeline provides a clear practical improvement over the plain backbone.

\subsection{Robustness Analysis}

To assess stability, we further performed three repeated runs on 120-task subsets. On the MBPP Full split, the observed improvement remained stable, with a 95\% confidence interval of $[0.0250, 0.1417]$. On the MBPP Sanitized split, the trend remained positive, with a 95\% confidence interval of $[-0.0194, 0.0722]$, although it crossed zero. Therefore, the robustness evidence is stronger on MBPP Full, while the Sanitized split should be interpreted more cautiously as showing a positive but less statistically stable trend under repeated subset evaluation.

\subsection{Contamination Analysis}

Finally, we conducted a contamination analysis using token-set Jaccard similarity:
\begin{equation}
    J(b, d) = \frac{|T(b) \cap T(d)|}{|T(b) \cup T(d)|},
\end{equation}
where $T(\cdot)$ denotes the token set of a benchmark instance or a training example. Across all training--benchmark pairs, both exact overlap and fuzzy overlap with $J \geq 0.8$ remained zero. This result reduces concerns that the observed gains are caused by prompt-level leakage or near-duplicate contamination between the training data and the evaluation benchmark.

\subsection{Discussion}

The results suggest that \textbf{Route} and \textbf{Align} help produce a candidate pool that is more favorable to correct solutions, but their effect becomes much more visible when combined with \textbf{Verify}. In other words, the main value of routing and aligned adaptation is not merely to improve a single direct output, but to increase the likelihood that correct or near-correct candidates appear in the generated pool. Verify can then exploit this improved pool and select a better final program.

This interpretation is consistent with the empirical pattern in Table~\ref{tab:ablation_mbpp}: the \textit{Route + Align} configuration without verification yields only limited gains, whereas the verified variants produce much larger improvements. Therefore, the advantage of RAV lies in the interaction among its components rather than in any single module alone. Route and Align reshape the candidate distribution, and Verify converts this advantage into stronger final pass@1 on MBPP.
\section{Conclusion}

This paper presented \textbf{RAV}, a modular framework for improving code generation functional correctness under a fixed backbone model. RAV combines three lightweight and complementary interventions: \textit{Route}, which applies task-aware prompt routing before generation; \textit{Align}, which reduces the mismatch between fine-tuning prompts and inference-time prompts through aligned LoRA adaptation; and \textit{Verify}, which selects final outputs using execution-based verification over generated candidates.

Experiments on MBPP demonstrated that the full RAV pipeline achieves the best performance among all evaluated configurations, reaching \textbf{0.8911} on MBPP Sanitized and \textbf{0.8520} on MBPP Full. Compared with the base model, this corresponds to improvements of \textbf{6.35} and \textbf{9.92} percentage points, respectively. The component-wise ablation further suggests that Route and Align are most effective when combined with Verify: prompt routing and aligned adaptation help produce a more favorable candidate pool, while verification converts this advantage into stronger final pass@1.

Additional analyses support the reliability of these findings. Repeated subset evaluation showed stronger stability on MBPP Full and a positive, though less statistically stable, trend on MBPP Sanitized. Moreover, contamination analysis found zero exact overlap and zero high-similarity overlap between training data and benchmark tasks, reducing concerns that the observed gains are due to prompt-level leakage.

Overall, the results indicate that functional correctness can be improved without modifying the backbone architecture itself. Instead, substantial gains can be obtained by coordinating \textit{how tasks are prompted}, \textit{how the model is adapted}, and \textit{how final outputs are selected}. As future work, a more complete factorial ablation, broader benchmark coverage, and stronger verification signals could further clarify the contribution and generality of the proposed framework.

\addtolength{\textheight}{-12cm}   
\bibliographystyle{IEEEtran}
\bibliography{ref}

\end{document}